# Evidence for High-Temperature Superconductivity in Doped Laser-Processed Sr-Ru-O

A. M. Gulian[1], K. S. Wood[2], D. Van Vechten[2], J. Claassen[2], R. J. Soulen, Jr.[2], S. Qadri[2], M. Osofsky[2], A. Lucarelli[3], G. Lüpke[3], G. R. Badalyan[4], V. S. Kuzanyan[4], A. S. Kuzanyan[4], and V. R. Nikoghosyan[4]

Abstract:

We have discovered that samples of a new material produced by special processing of crystals of $Sr_2RuO_4$ (which is known to be a triplet superconductor with $T_c$ values ~1.0-1.5K) exhibit signatures of superconductivity (zero DC resistance and expulsion of magnetic flux) at temperatures exceeding 200K. The special processing includes deposition of a silver coating and laser micromachining; Ag doping and enhanced oxygen are observed in the resultant surface layer. The transition, whether measured resistively or by magnetic field expulsion, is broad. When the transition is registered by resistive methods, the critical temperature is markedly reduced when the measuring current is increased. The resistance disappears by about 190K. The highest value of $T_c$ registered by magneto-optical visualization is about 220K and even higher values (up to 250K) are indicated from the SQUID-magnetometer measurements.

[1] Physics Art Frontiers, Ashton, Maryland, 20861-9747
[2] Naval Research Laboratory, Washington DC, 20375-0001
[3] The College of William and Mary, Virginia, 23185
[4] Physics Research Institute, Ashtarak, 378410, Armenia

1. Introduction.
Laser ablation is one of the non-traditional methods of materials processing that may substantially alter the physical properties of samples. In addition to removing atoms from the surface, it may leave behind a recrystallized surface layer of altered composition and properties. In this work we applied this method to highly perfect crystals of the triplet superconductor $Sr_2RuO_4$ . SEM images show the resulting surface to be very lumpy and of non-uniform reflectivity. EDAX indicates the average composition to be Ag-doped $Sr_2RuO_{6\pm x}$ (x<1). This new material reveals dramatic features. As we deduced from our measurements and analysis, the top "crust" layer, most affected by the laser processing, is a superconductor with a transition in the 200-250K range. The superconducting layer thickness is estimated to be 4-5 μm. The deeper layers (total thickness of the samples is 25 μm) have also been affected, and demonstrate pronounced ferromagnetism at temperatures as high as room temperature. This is itself also interesting since Sr-Ru-O systems never before demonstrated ferromagnetism at temperatures higher than 150K. In this article we focus on superconducting properties of the samples, the findings on ferromagnetic behavior will be reported separately.

2. Sample preparation.
All samples were manufactured from a single strontium ruthenate crystal (#C148) measured to have $T_c$ = 1.13 K and grown in accordance with current standard procedures. Using the natural anisotropy of such crystals, many slices (perpendicular to the *c*-axis) were peeled from the initial crystal. The slices were then polished on opposing *ab*-plane faces into free-standing plates about 25 μm thick and 1-3 mm in lateral dimensions. Each plate was then glued onto a glass substrate for further processing that began with deposition of  ~ 3000Å thick Ag layer on top of the glass substrate and the mounted $Sr_2RuO_4$ slice. Then laser micromachining was used to separate the internal part of the sample from the edges and define four conductive legs extending from the outer area to within 10 microns of the center of the plate. Finally, laser ablation was used to remove the silver from the central part of the crystal. The goal was to make a small region of the center of the plate available for R(T)-measurements using standard four probe techniques with the probes made of the same material. (Our goal was to obtain triplet superconductivity in a small, single-domain volume, both to explore possible direct detection of chiral currents and for other exploratory purposes having to do with quantum computing.  These aims are relevant only as explanation for the geometry that was produced.) The final geometry of the samples is explained in Figure 1.
      Special protection from air appears not to be necessary. All samples have been stored in air without any desiccators or surface passivation layer. Moreover, sample #1 displayed consistent signatures of superconductivity at times separated by a year of such storage.

3. Sample composition.
We performed X-ray (EDAX) microanalysis of the laser-micromachined surface (Table 1). A test revealed that this composition persists to a depth of not more than 2 μm from the surface of the sample. As follows from this Table, the Sr/Ru ratio is about 2, so that we are far from the so called Ruddlesden-Popper series with composition $Sr_{n+1}Ru_nO_{3n+1}$ which reveals ferromagnetism at n>1, with the highest onset temperature ~150K at n = ∞ ($SrRuO_3$).

**Table 1**

| Element | SEM Microprobes (in at%) | | | Average |
|---|---|---|---|---|
| Strontium | 21.50 | 23.36 | 22.53 | 22.46 |
| Ruthenium | 10.43 | 11.51 | 10.91 | 10.95 |
| Oxygen | 65.98 | 64.75 | 65.86 | 65.53 |
| Silver | 2.09 | 0.38 | 0.70 | 1.06 |

The presence of excess oxygen is the most striking feature in Table 1. There is also some presence of Ag. It is clear that both excess oxygen and silver doping are introduced from the free surface. Thus, at sufficient depth into the affected layer the composition should correspond to the mother-substance with $O_4$ rather than $O_6$, and no Ag. Intermediate compositions are obviously present, and will be seen below to have properties such as ferromagnetism, paramagnetism and superconductivity.

We measured a diffraction scan of laser-processed sample 2 (the second sample to show a resistivity transition) and a polished but otherwise unmodified sample of the parent 214 material. In the plot of intensity vs. Bragg angle, the parent sample revealed seven major peaks catalogued for the material, plus some weaker peaks associated with a superlattice. Surprisingly, the scan from the laser processed sample shows a higher crystalline quality than the parent. The 214 features become stronger, with the most dramatic growth in intensity occurring in features representing shorter d spacing. The superlattice peaks shift to smaller Bragg angles indicating that the superlattice has expanded. Additional sharp features appear which we are still analyzing. The most important aspect of these is a new feature indicative of very large superlattice spacing not present in the parent material. To recapitulate, the analysis confirms that the peculiar processing has not resulted in chaos, but has instead given new crystalline substances, such as would be consistent with superconductivity.

4. Resistivity measurements.
Multiple resistive transitions were recorded on two samples with two different laboratory setups. Curves for the best characterized sample are shown in Figure 2. Sample #2 demonstrated the same type of transition, until contact through the central small bridge was lost. Figure 3 documents the suppression of the onset temperature caused by increasing the measuring current. This figure shows the raw data as the sample cooled. At times prior to the transition the slowly-alternating sign of the measuring current creates a rectangular periodic voltage structure. At the transition, this square wave pattern disappears and is replaced by the system noise and temperature dependent offset voltage associated with the leads from room temperature (thermoelectricity related). Comparing the height of periodic voltage above the transition with the noise amplitude, we deduce that the resistance dropped by a factor of at least 6000. Very roughly our experimental configuration (Figure 1, extreme right) can be considered as a bridge with approximate sizes: length $l \sim 10\mu m$, width $w \sim 10\mu m$, and height $h \sim 25\mu m$. Since $\rho = Rwh/l$, our measured $R_{300} \sim 0.02\Omega$ yields $\rho_{300} \sim 50\ \mu\Omega cm$; this is a good accordance with the known value $\rho_{300} \sim 100\ \mu\Omega\ cm$ for $Sr_2RuO_4$ in view of crude model for the bridge geometry. Below the transition $\rho < 0.01\ \mu\Omega\ cm$ if the current flows through the whole crossection of the bridge, or $\rho < 0.001\ \mu\Omega\ cm$ if the current flows through the micron-order thick surface "crust". These values for conductivity cannot be attributed to any normal metal. Moreover, one cannot explain such a drastic change in resistance by a shunting effect: a factor of 6000 of R-reduction would require the shunt, if made of the same material, to have a 6000 times larger cross section than the central connector, more than filling all the trenches. But transmission light microscopy with illumination from underneath the sample shows no such material in the trenches. The hypothetical possibility that optically invisible nanobridges exist and are responsible for the transition is possible only if they are themselves high-temperature superconductors. But even if they filled the trench, their volume would still be too small to produce the amplitude of the diamagnetic signal discussed below.

5. Magnetization measurements by SQUID-magnetometer.

To trace the Meissner effect we undertook SQUID-magnetometer (Quantum Design MPMS model) measurements. The results are shown in Figure 4-6. In plotting curves in Figure 4a we show raw data for magnetic moment vs. temperature, without any corrections, and we also show the same curve for the mother substance mounted on its substrate from a reference sample. The resultant curves are not as simple as for YBCO and comparison yields a straightforward suggestion that some positive additional contribution (ferromagnetic or superparamagnetic) is present. Figure 4c demonstrates that ferromagnetism exists in the laser-processed samples at 300K. Superposition of diamagnetic and ferromagnetic responses causes the curves in Figure 4a to start at positive values and changes the shape of the H=100 Oe curves compared to H=10 Oe curves. Indeed, the diamagnetism causes change in sign of the derivative with respect to T on the zero field cooled curves of Figure 4a occurs at ~250K at H=10 Oe, and ~70K at 100 Oe. At low T, the diamagnetic signal exceeds the ferromagnetic one. The ferromagnetic features present at 300K in the magnetization of the laser-processed sample (Figure 4c) are absent in the unprocessed reference sample. The curves at 300K have also a large-scale diamagnetic background because of the substrate (glass, 1mm thick) contribution – it can be easily subtracted, as well as the paramagnetism seen in the reference and noticeable in Figures 5e and 5f. When the 300K curve is subtracted from the raw data (Figures 5a,c, and e) there results a signature similar to a distorted superconducting butterfly (Figures 5b,d, and f). In Figure 6 we demonstrate that a classical "butterfly" signature can be generated by making two further subtractions, one of them a paramagnetic component and ferromagnetic component (representing changes from 300K of both the paramagnetic and ferromagnetic terms). Careful attention to the starting portion of the M(H) loops indicates that the Meissner effect disappears at about 250K.

6. Magneto-optical imaging measurements.

To better understand the laser-processed samples, magneto-optical (MO) imaging was employed. The experimental setup (Figure 7a) consisted of a custom assembled polarizing microscope built around a continuous-flow cryostat. On top of the specimen we mounted an epitaxially grown bismuth-substituted ferrite-garnet film with in-plane magnetization and illuminated the surface with polarized monochromatic light. The reflected off the bottom of the iron garnet photons experience a pronounced Faraday rotation of the polarization, the magnitude of which is proportional to the local magnetic field. Therefore spatial distribution of the polarization of the light acts to image magnetic flux lines. Figures 7b and c indicate the change in the response to a YBCO sample at the transition. MO images of the laser-processed strontium ruthenate sample were taken at various temperatures from 300K down to 60 K, with an applied external field of about 10 mT parallel to the c-axis. Clear signatures of magnetic flux lines start to appear at 220

K as bright small spots (Figure 8c) in contrast with the homogeneous background (Figure 8b), and are aligned in a narrow segment on the central-right part of Figure 8d. At cooling down these features grow in brightness and in number indicating that magnetic field lines are accumulating in an inhomogeneous way in specific spots of the sample surface. A comparison between Figure 8a and Figure 8d is shown in Figure 8f and reveals that the path formed by magnetic field lines mainly follows the crossed trenches created in the sample by laser ablation. Some of the field lines align along the perimeter of the doped area corresponding to the black rectangle in Figure 8a or equivalently the darker part shown in the SEM measurements in the center of Figure1. This effect is even clearer at lower temperature where in the Figure 8e, measured at 60 K, the border of the lower left quarter of the doped rectangle is marked by continuous dotted lines. As shown in Figure 8f, at this temperature magnetic field lines are well formed and their alignment along the trenches appears to follow the local roughness of the walls. The accumulation of magnetic field lines identified as bright spots by the MO imaging reveals the presence of local magnetic effects that could, in the absence of further information, be attributed to diamagnetism or ferromagnetism. However, correlating this MO information with the SQUID information shows that the MO images change at the temperatures where diamagnetism manifests itself in cooling, and where ferromagnetism is roughly constant; hence the preferred interpretation is that the fine structures along the trenches represent diamagnetism. The alternative hypothesis that the bright spots are artifacts caused by the surface roughness and mechanical stress was excluded by the absence of MO response to intentionally roughened samples.

7. Discussion.

We have obtained compelling evidence that our synthesized material is a very serious candidate for a new very high-$T_c$ superconductor. As Figure 6 demonstrates it is a type II superconductor. At this point it is not clear whether oxygen doping, silver doping, or lattice distortion play critical roles in the superconductivity. In addition, one cannot neglect strong distortion of the crystalline lattice. Since the critical temperature is high, the coherence length is most likely very short, and there is no preclusion of pairing with orbital angular momentum greater than zero. For the same reasons, triplet state superconductivity cannot be ruled out.

For quantitative comparison, the YBCO film used as a control was made of approximately the same surface area as the laser-processed sample. Its magnetization amplitude is a factor of 20 smaller than that of our sample #1. Meanwhile, the MO imagery shows that trenches are also contributing. Their surface area (including 25μm high double-walls) is quite comparable to the plane top surface. This means that the depth of the active superconducting volume (the laser-processed crust) is about ten times more than that of the YBCO-film, i.e., about 4μm. This tells that the MO imaging was detecting the superconducting phase, which became detectable at 220K. The ferromagnetic phase, which was present at room temperatures with the same spatial contours, was not seeable because it was deeper (microns away from the surface) which precludes sharp imaging in the MO technique. Most likely the ferromagnetic phase is located intermediately between the superconducting surface layer (more enriched by oxygen) and the mother-substance.

It is clearly necessary to find out more about the role of the deposited metallic layer, switching from Ag to another metal or excluding metals completely. Another task is to reproduce the effective composition by methods that yield homogeneous samples. While work is ongoing in these directions, it is necessary to mention that traditional methods of synthesizing samples may not work if the equilibrium phase diagram precludes the required composition or/and the (as yet unknown) structure we have obtained. In our approach, it is relatively simple to vary the composition of overlying material and laser pulse characteristics and then determine the extent of implantation effects.

8. Conclusions.

We have presented a case for high temperature superconductivity based upon resistivity, SQUID measurements, magneto-optic imagery, and X-ray analysis. This paper is a summary of our results and a better version will be offered soon. It is still possible that an alternative interpretation could be found that fits all these results, but we have not succeeded in constructing one. It is not enough to reject the resistivity and magnetic moment and imagery results individually; there must be an integrated physical scenario for how laser processing could have produced the entire set of results. Accordingly we feel it is appropriate to present the information that we have.

The following recapitulates the main points, according to the interpretation we have defended.
a) Starting from $Sr_2RuO_4$ single crystals and using laser processing, we reproducibly manufactured new material(s) with enhanced oxygen content, and which collectively demonstrate both superconductivity and ferromagnetism.
b) Multiple resistance measurements were carried on two of these samples, and the results are consistent with a superconducting transition. No short circuiting or current lead disconnection can be responsible for these observations.
c) Magnetization measurements reveal the existence of a (ferro)-magnetic phase in the sample volume, as well as a diamagnetic phase. The latter reveals the classical butterfly signature of type II superconductivity after the ferromagnetic phase signal is subtracted.
d) Diamagnetic signatures of superconductivity disappear at about 250K, while the ferromagnetic component still exists at room temperature.
e) Appearance of the (dia)-magnetic component at temperatures above 200K is clearly visible in the MO technique. This measurement, combined with the magnetization measurements, allows us to deduce that the ferromagnetic phase is probably located in the deeper layers, deeper than the superconducting layer and above the mother-crystal.
f) Though we did not perform quenching of superconductivity by a magnetic field, we succeeded in quenching of superconductivity by current.
g) The estimated value of the critical current, as well as the amplitude of the diamagnetic response are within the reasonable ranges.

We are grateful to Y. Maeno and H. Yaguchi for providing us $Sr_2RuO_4$ crystals and many useful discussions. AMG is grateful to H.-D. Wu, V. Thai, B. Rinker, B. Phlips, J. Kurfess and J. Horwitz for their help and support. VRN is grateful to S. Petrosyan and A. Melkonyan for the assistance in the lab work. NRL researchers and AMG would like to express their appreciation of the support and encouragement of H. Gursky and D. Gubser. KSW wishes to acknowledge that part of his contribution was done while he was at the Aspen Center for Physics. This work at NRL was supported in part by Navy basic research funding, and by the ONR Grant

N0001403WX20850. The work by DVV is supported by the ONR ROPO program. The work at CWM was supported in part by DOE grant DE-FG02-04ER46127.

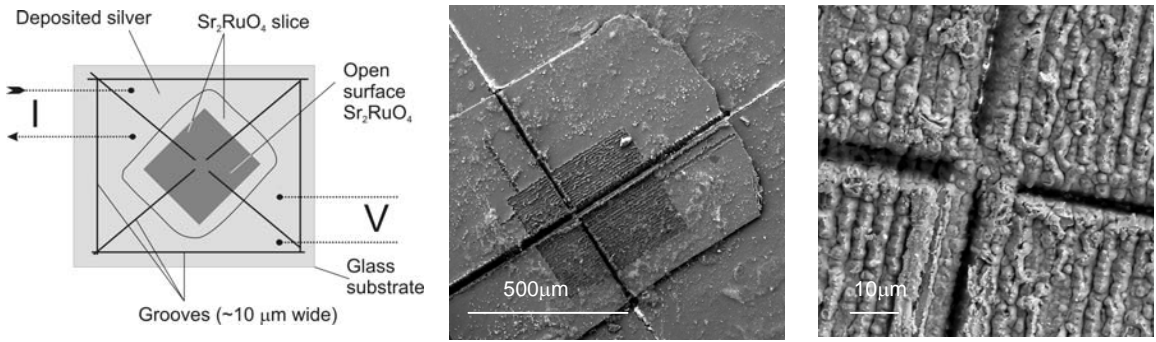

Figure 1. *(left)*-schematic of the sample and the measurement connections, *(center)*-SEM of the processed sample #1, *(right)*-central part of sample #2.

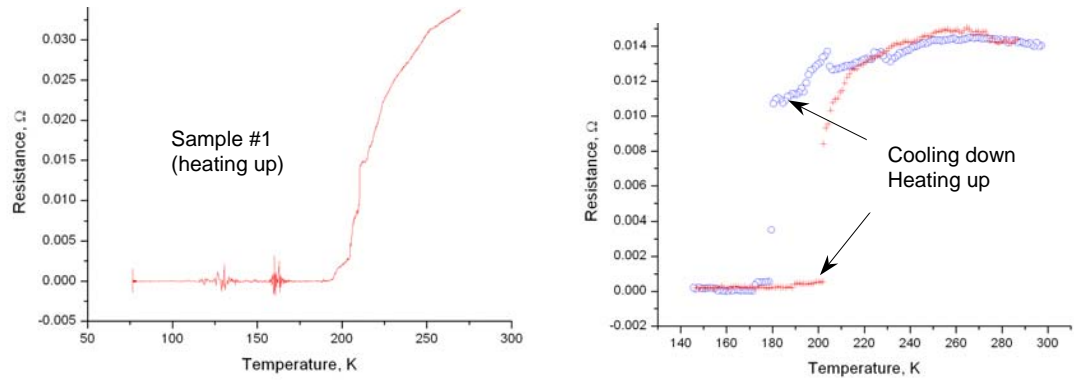

Figure 2. Resistive measurements with sample #1 performed in two different experimental setups. Different contact pads were chosen in the two experiments and thus the resistance values of the bridge at 300K are not identical. The left curve was measured at 1 mA current, whereas the right curve was measured at 10mA. The large current in the latter case almost surely produced the hysteresis.

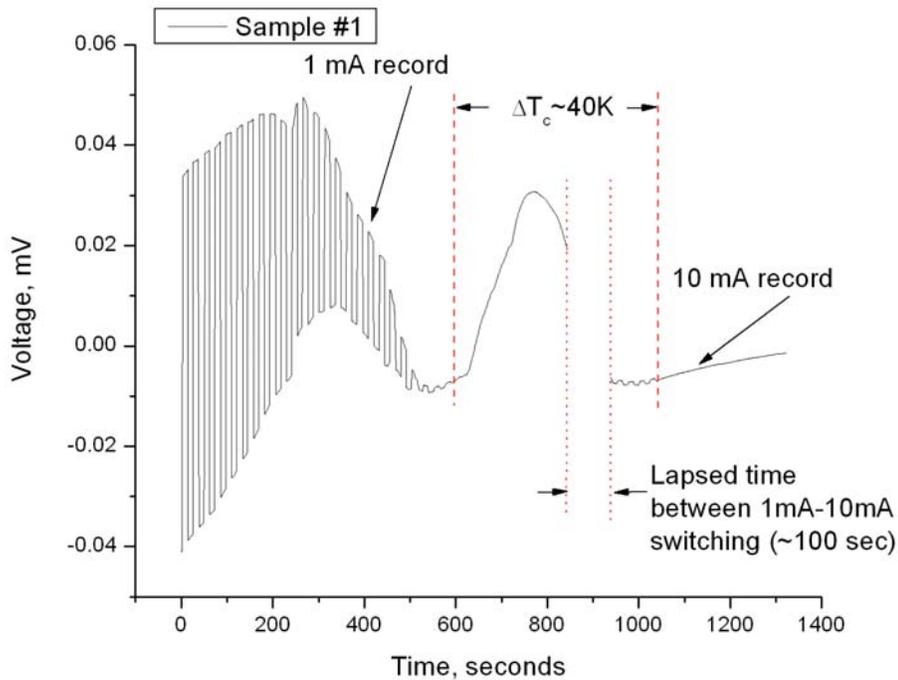

Figure 3. Suppression of transition by applied current.
The voltage vs. time during the cool down was measured with a DC current of alternating polarity passing through the sample. Accordingly, voltage jumped up and down, so that the height of the vertical lines is the measure of the sample resistance. At 1 mA of registering current the transition occurred at t=600 sec, while switching to 10 mA of registering current caused a return to the resistive state (represented by the ripples near t=930-1050sec) followed by a new transition at t=1050 sec, corresponding to $\Delta T_c$~30-40K. (Note that V(t) signal, which is intentionally preserved in its original form in the figure, also contains a T-dependent thermoelectric DC component because of the cryostat wiring.)

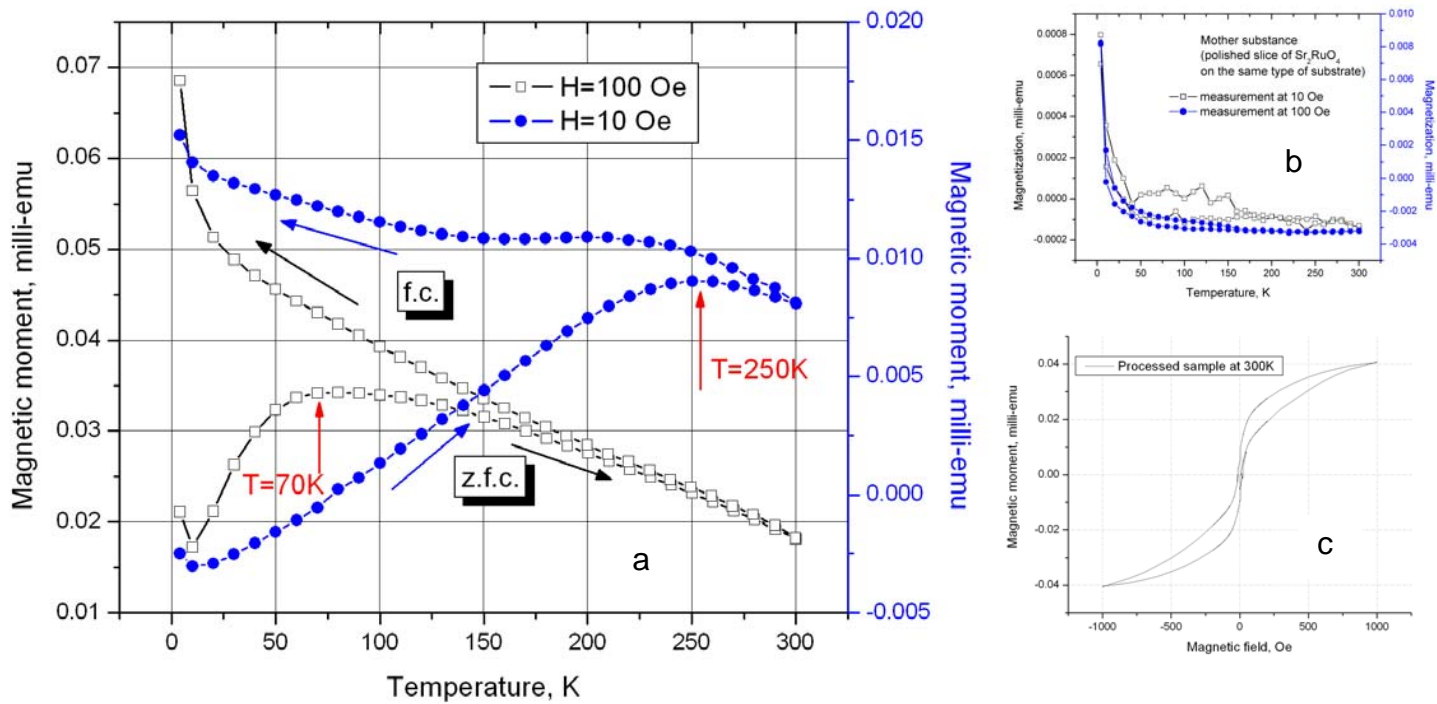

Figure 4. Magnetization vs. temperature of the laser-processed *(left, a)* and unprocessed *(control sample, up-right, b)*. For each value of H there is a field cooled (f.c.) and a zero-field-cooled (z.f.c.) part. Panel a reveals that ferromagnetism is present at room temperatures in the processed samples. This is further substantiated in panel c *(bottom right)*.

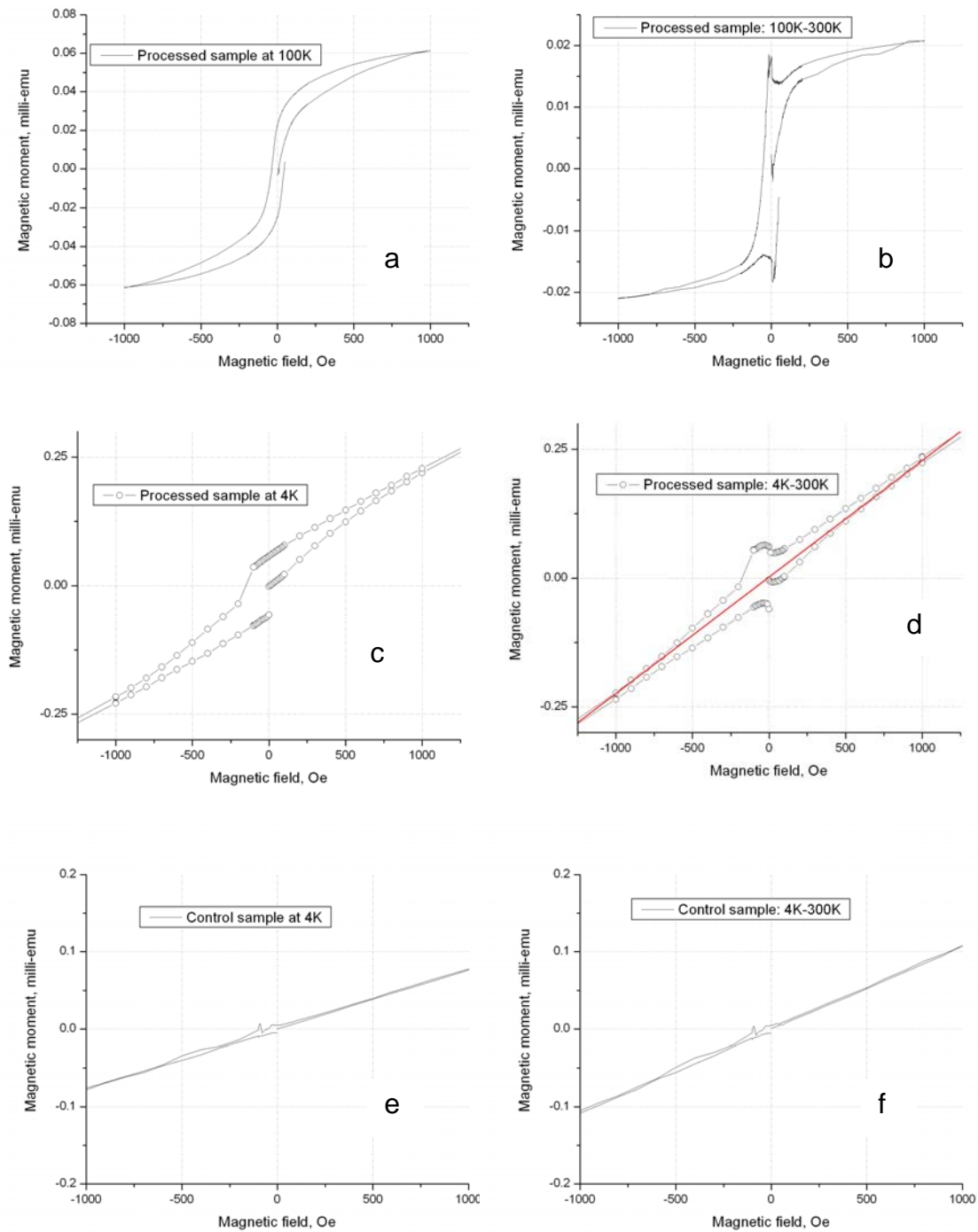

Figure 5. Magnetic moment *vs.* magnetic field, laser processing and temperature:
a) processed sample at 100K;
b) processed sample at 100K, but with magnetic moment background (Figure 4c) at 300K subtracted;
c) processed sample at 4K;
d) processed sample at 4K, but with magnetic moment background at 300K subtracted; a red line indicates the paramagnetic background subtracted to produce Figure 6a;
e) control sample (unprocessed) at 4K;
f) control sample (unprocessed) at 4K, but with magnetic moment background at 300K subtracted.

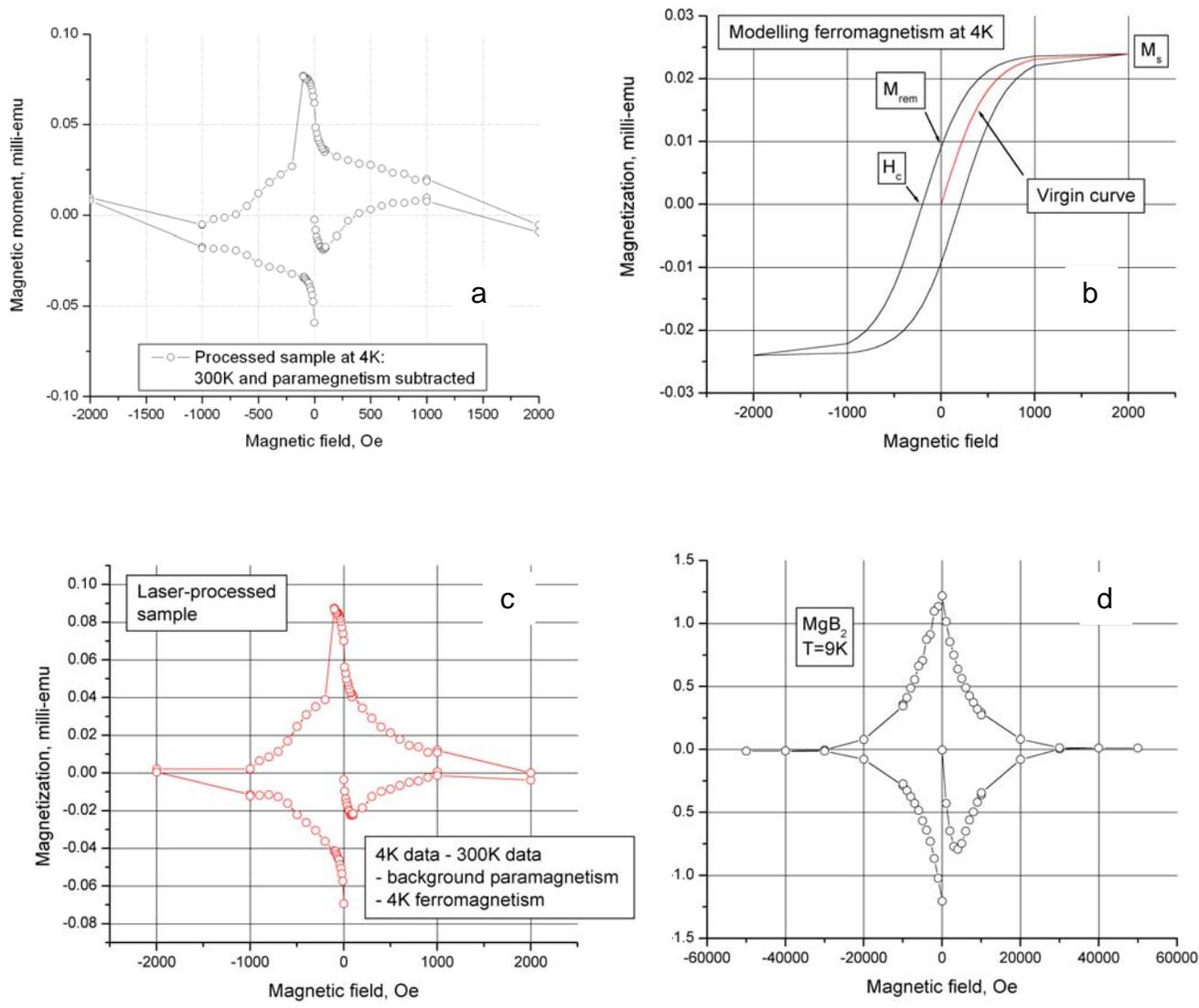

Figure 6. Magnetic moment of sample #1 compared to a classical "butterfly" signature of superconductivity ($MgB_2$) which we measured with the same apparatus. Subtraction of paramagnetic background (red line in Figure 5d) yields panel 6a. The distortion compared to a classic butterfly is due to the change in the ferromagnetic signal from room temperature to 4K, modeled as in panel 6b. Subtracting it from 6a yields the more perfect butterfly in panel 6c, which is quite congruent to a traditional one (6d).

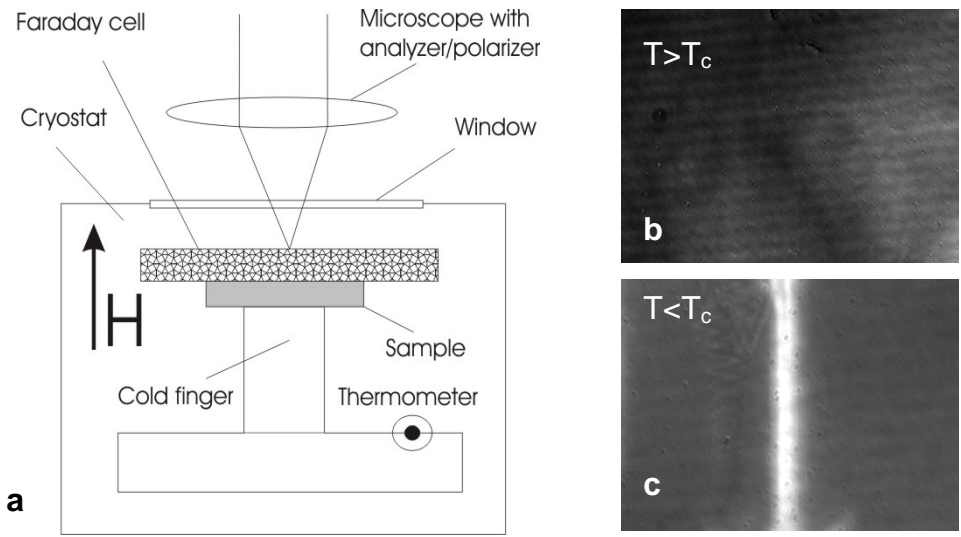

Figure 7. a - Experimental set up for visualization of magnetic flux distribution via the Faraday effect. The structure labeled as "Faraday cell" includes the Bi-substitute ferrite garnet film described in the text. (The thermometer and window geometry are such that the sample cannot be colder than the thermometer.) 7b and c: visualization of a scratch in a YBCO film; b above Tc and c below Tc, to show how the MOI signature disappears at the transition.

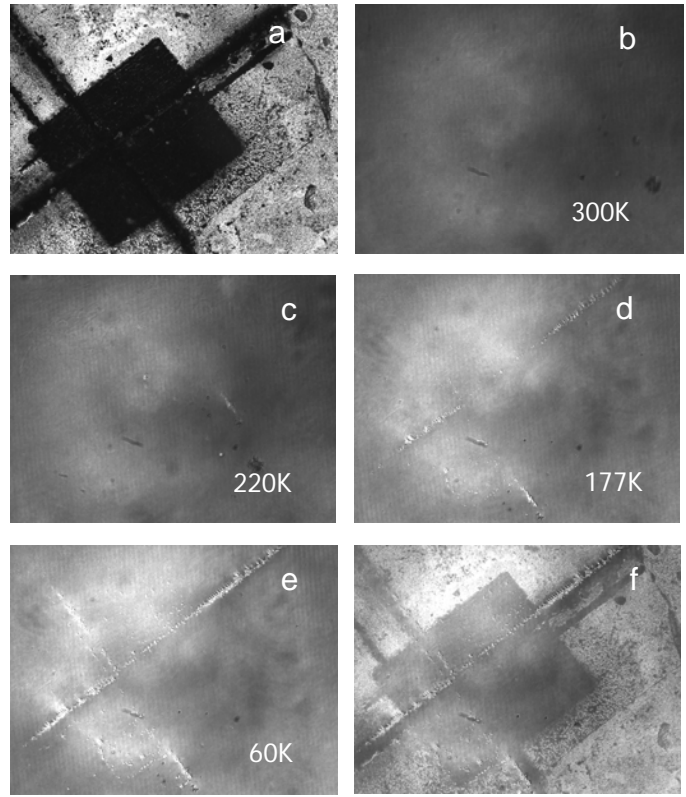

Figure 8: Magneto-optical images (MOI) of sample #1.
Panel a) shows the appearance of the sample at the same magnification as the other photos and without the overlaid indicator material; b) indicator material is overlaid: no-signal background of MOI; c) first temperature at which evidence of localized flux was noticed; d) trenches well decorated with localized flux and fingers forming, perimeter of processed area clearly visible in bottom left quadrant; e) completely developed image; f) panels a) and e) superimposed.